# Development of System Architecture for E-Government Cloud Platforms

Margulan Aubakirov
JSC National Information Technologies
Astana, Kazakhstan

Evgeny Nikulchev
Moscow Technological Institute
Moscow, Russia

*Abstract*—Requirements and criteria for selection of cloud platform and platform visualization are stated by which optimal cloud products will be chosen for the Republic of Kazakhstan e-Government considering quality-price ratio, and also the framework of information and communication architecture will be introduced.

*Keywords—cloud technologies; outsourcing; Kazakhstan; cloud platform; e-Government*

I. INTRODUCTION

Cloud technologies is an effective model for reduction of aggregated value of information systems ownership due to resources incorporation to shared pool (for example, computer capacities, data storage systems, channel capacity and memory) from which resources can be immediately allocated and deployed in accordance with changes made to requirements.

The cloud services provide an opportunity for users to keep data (for example, pictures or e-mail messages), use software (for example, social media, video and audio files, games, etc.). For companies, including government agencies, cloud services can be applied as substitutes to internal data centers and departments in charge of ICT [1]. Companies not having equity contributions to the creation of information infrastructure can suggest services and work to their future customers. In general, cloud technologies embody a further industrialization (standardization, scaling-up, major distribution) just like provision of electrical power by electric supply stations to end users. Due to a standardized interface, the absence of necessity to solve issues of data centers creation, launch and safety provision and servicing of a variety of users it is possible to achieve an effect from the scale. In this regard, cloud technologies distribution on the national level enables achieving an optimal expenditure level [2]. Economic surveys results acknowledge the importance of cloud technologies application and forecast their distribution throughout the world [3].

Application of cloud technologies allow decreasing capital and current operating expenses and increase the level of equipment usage, which currently amounts to 10% in public sector. It means that 90% of actually acquired equipment in public sector is on downtime, which in its turn means its inefficient use [4].

Thus, the common objective of a new government agencies' informatization service model development is reduction of expenses for IT-resources management; reduction of costs for IT-personnel, optimization of budget costs on procurement of IT-equipment, implementation of single pricing policy, increase of servicing quality; enhancement of government agencies' IT infrastructure and information safety, reduction of risks regarding data loss and corruption [4]. Above-mentioned conditions can boost the percent of Kazakhstani participation in services procured by government agencies and provided by cloud platform operator.

In this regard, cloud computing is designated as crucial technology according to the result of technological forecast executed as part of Framework for Innovative Development of the Republic of Kazakhstan until 2020 in the field of information and communication technologies.

One of the mechanisms for efficiency enhancement of information technologies application in government agencies is the implementation of a new informatization model based on migration to the use of cloud computing, ICT-outsourcing and orders consolidation.

The results of the technology implementation are as follows: budget consolidation and saving, government agencies' business-processes efficiency.

As part of G-Cloud realization, on the 1st phase of the project, it is planned to deliver IaaS (the service for the provision of the virtual machine and service for allocation of virtual space for backup and storage) and SaaS (the service of e-workflow, e-postal service).

II. ANALYSIS OF CONDITIONS FOR CREATION OF G-CLOUD PUBLIC PLATFORM IN THE REPUBLIC OF KAZAKHSTAN

One of the mechanisms for efficiency enhancement of information technologies application in government agencies is implementation of a new informatization model based on migration to application of cloud computing. Proposed informatization (service) model of government agencies implies the consolidation of government agencies' IT-infrastructure, provision of a number of IT-services in accordance with principles of "cloud computing", including such services as e-mails, e-workflow and others.

The project implementation will address the following issues.

Today, development of information technologies in government agencies is characterized by the following issues:

- Procurement and maintenance of IT-equipment is not government agencies specialty.

- Unequal level of required IT-infrastructure provision and maintenance.
- Partial non-compliance with requirements of life sustenance systems backup.
- Impossibility to track the level of equipment load and maximum effective use.

Annually, government agencies provide budget for maintenance and development of existing information systems, infrastructure required for systems operation and procurement of servers, network and cross connect equipment, etc.

The first point is that costs on server equipment maintenance continue to rise by two-three times. In case if some information system requires more resources, additional purchase of necessary equipment or server`s complete replacement will be made. The process of server replacement (beginning from procurement and finishing with the system setup) might last for more than a month. In other words, the problem of server farm modernization will only become worse and lead to the slowdown of government agencies' informatization process, which means more frequent equipment failures, unsatisfactory provision of services and citizens dissatisfaction. Development of public cloud platform will help to solve problems with inefficient use of equipment, provide a required safety level and information systems efficiency.

Secondly, each government agency procured or sold information systems for automation of certain functions in order to boost efficiency. Therefore, budget funds were consistently provided for automation of government agencies standard functions. Besides, government agencies provide budget for maintenance and development of existing information systems, infrastructure required for systems operation and procurement of servers, network and cross connect equipment, etc. on an annual basis. All these lead to irrational spending of budget funds as in practice funds are spent several times for the same purposes by different agencies.

Likewise, when retail software is acquired for functions automation, agencies face the problem when a product doesn`t meet all the requirements. In this case, agencies have to provide additional budget resources for upgrading/development of such solutions. Automation of such standard functions inside each agency are partial as agencies procure /sell various solutions for automation of certain activities, which in its turn, inevitably leads to a large number of information systems inside an agency, each of which automates only one function. In the long run, agencies will have to address the problem of various information systems management, which considerably increases costs.

Annually, government agencies spend considerable budget funds on procurement of computer equipment (including servers), salaries and administrative expenditures of information and communication units of government agencies.

Also, there is imbalance in government agencies information and communication assets (availability of excessive assets in government agencies executing minute volume of functions and providing narrow spectrum of public services and lack of computation capacities in government agencies executing a large volume of functions and providing a wide spectrum of public services). The most part of procured technical means are not used in full extent and rapidly become obsolete because of technical progress in the field of informatization.

Government agencies performance has general and similar components. Today, each government agency solves the issue of informatization on their own, acquiring IC-equipment and software, creating data centers and automating public services. This includes budgeting and its approval for inspection of government agencies' IC-infrastructure, consulting services, feasibility study, software development, equipment procurement, licensing, maintenance and also, execution of tenders and contracting.

In many government agencies and quasi-public sector unlicensed software is used, which is a violation of international liabilities of the Republic of Kazakhstan in the field of intellectual property protection. To solve these problems it is planned to migrate on cloud computing technology based on virtualization one. Having cloud computing technology, a user gets full-featured virtual server that equals physical one.

As the result, there would be no necessity for government agencies to acquire physical equipment, service it and bear expenditures on salaries for personnel.

### III. REQUIREMENTS FOR CLOUD PLATFORM

Information and communication platform (hereinafter – ICP or G-Cloud) is hardware and software package intended to provide services to government agencies in the field of informatization with application of "cloud" technologies. Information and communication service (IC service) is an assembly of services for rent and allocation of computing resources, provision of software and equipment and also, communication services via which the stated services operate.

As part of G-Cloud implementation, on the 1st phase of the project it is planned to deliver IaaS (the service for provision of virtual machine and service for allocation of virtual space for backup and storage) and SaaS (the service of e-workflow, e-postal service).

Fig. 1. demonstrates resources management by client and managing company for every type of main services.

The main requirement specified for G-cloud platform is that it should be based on leading solution in the field regarding development of virtual and cloud infrastructures.

Cloud platform should provide a quick transfer of existing government agencies data centers to computing cloud. The platform should offer such possibilities as high availability, data recovery, "hot" migration, "hot" resources inclusion, resilience, automated distribution of resources, virtual hub for serial ports and store API-interfaces for integration of sets and maintenance of alternative I/O ways.

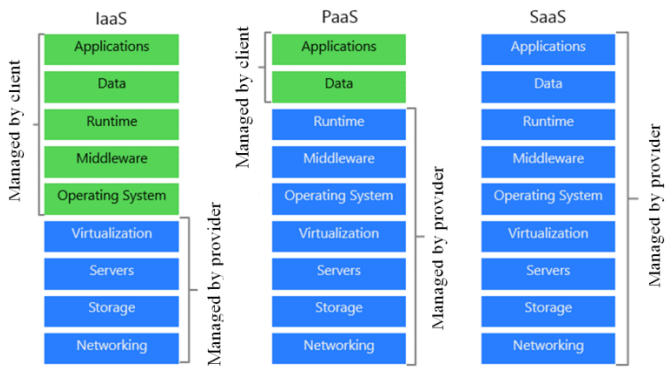

Fig. 1. Levels of resources management for various types of services

Application of cloud platform should provide a consolidation of existing infrastructure and optimization of government agencies' IT-equipment, ensure flexible scaling and reduce costs for solutions regarding provision of services operation continuity and emergency recovery.

The platform should ensure the possibility of functionality expansion through installation of additional modules and possibility to create cloud infrastructure (private or public computing cloud) with several renters due to resources incorporation in virtual data centers. Afterwards, the centers are offered to users via web-portals and software interfaces as fully automated and catalogued services.

The platform should support three-landscape approach for systems implementation:

*Development environment* is an environment which has a complex of programs with functionality required for systems development [6]. "Development environment" concept can be understood in various ways. For instance, a programmer`s main tool is integrated development environment, which has all the tools for creation of codes, compilation and so on. On the higher level, development environment is an adjusted environment together with development server, data, data base processor and other required tools needed for examination and testing of developing system`s components performance.

*Test environment* is an important link between development environment and real production environment. It consists of equipment (servers, operating computer (s), etc.) and components of logical level (server operating system, client operating system, database server, client user interface, web-browser (explorer) and other software [7]). The environment encompasses client components as well as server ones and uses the same versions of software that are located on clients, i.e. similar to production environment if possible. Otherwise, during the implementation process it might turn out that the system doesn`t work and needs readjustment. In test environment performance tests and efficiency tests are executed together with tests on systems upgrade and error control. Also, user acceptance tests can be performed. There is a thumb-rule according to which the whole test environment is separated from production environment and that all updates and adjustments are controlled in test environment and only afterwards are installed to production environment. Test environment is also suitable for clients training, which guarantees that during the process data, for instance, won`t be corrupted.

*Production environment* is an environment in which actual work is executed, i.e. activities executed by a company every day. Similar to test environment, production environment consists of complete software and hardware package.

Key requirements:

*1)* Support of open standards for storage and distribution of virtual machines (OVF);
*2)* Open software interface for integration with external systems;
*3)* Support of possibility to use cloning for deployment of virtual machines and applications;
*4)* Support of virtual machines bound copies constructed on the basis of "golden" patterns in order to save disk space of storage system and optimize applications deployment;
*5)* Built-in features of virtual infrastructure network security;
*6)* Support of infrastructure services catalogues with possibility of services publication in the catalogue by users;
*7)* Logical partitioning of resources pools provided by virtualization platform on virtual computing data centers with fixed service quality;
*8)* Support of operation with various organizations: isolation of virtual resources, independent LDAP authentication;
*9)* Self-service portal for users and administrators access;
*10)* Enhanced users capabilities on independent management of organizations infrastructure (computing resources, storage resources and local network resources management);
*11)* Support of virtual distributed switch;
*12)* Integration with solutions on provision of network safety;
*13)* Possibility to create protected network circuit for data exchange between various cloud infrastructures.

### IV. CLOUD PLATFORM FUNCTIONAL STRUCTURE

In this section we`ll focus on cloud platform functional structure scheme for e-government agencies', its subsystems and their functionality. Fig. 2 demonstrates the scheme of infrastructural division.

"ICP for Internet" and "ICP for government agencies' Intranet" platforms are functionally identical and contain the following subsystems allocated by purpose and functionality. Pic. 3 demonstrates the scheme of ICP functional structure.

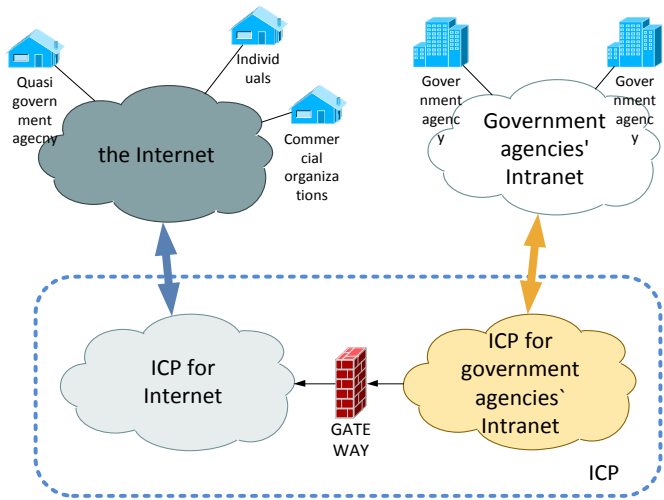

Fig. 2. Scheme for infrasrucutral division (G-cloud)

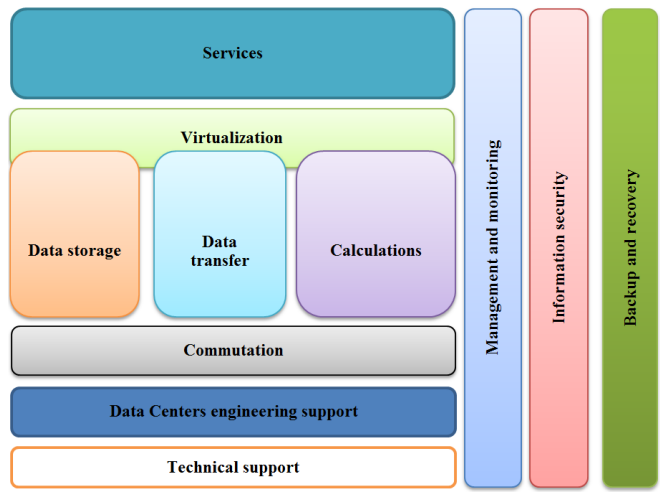

Fig. 3. Scheme of ICP functional structure

*Subsystem of services* provides key services (cloud services) to the system users.

*Subsystem of resources virtualization* implements software virtualization of physical computing resources.

*Subsystem of computing platform* provides the platform of unified physical servers.

*Subsystem of data transfer* provides traffic transfer in virtual environment.

*Subsystem of data storage* provides equipment for data storage and distribution. File and block access to data.

*Subsystem of commutation* provides physical computer network.

*Subsystem of data centers* engineering assistance provides data centers engineering subsystems.

*Subsystem of management and monitoring* provides subsystems management and monitoring functionality.

*Subsystem of information safety* provision provides the required information safety package.

*Subsystem of backup and recovery* provides the required functionality for data backup and recovery.

*Subsystem of technical assistance* provides organizational and technical support of the system operation.

V. SERVICES SUBSYSTEM

Services subsystem provides users with software services, classification of which match below-mentioned criteria.

**IaaS** (infrastructure as a service). Rent of virtual capacities in cloud. Service model which provides users with virtualized technological infrastructure using which it is possible to deploy and execute software, including operating systems and server applications. Control and management over major physical and virtual cloud infrastructures, including networks, servers, types of operating systems in use, storage systems is executed by cloud provider [8].

**PaaS** (platform as a service). Ret of platforms for developers. Service model which provides users with environment for code deployment and execution, creation or acquisition of applications on cloud infrastructure with application of tools and programing languages supported by the platform with integrated service of e-Gov infrastructure. Management and control over major physical and virtual cloud infrastructure, including networks, servers and operating systems are executed by cloud provider except developed and installed applications and also, if possible control is provided f environment (platform) configuration settings [8].

**SaaS** (software (application) as a service). Service model providing users with access to various applications operating in cloud infrastructure. Applications can be of various types and available from all devices with different operating systems. Users' access to applications can be provided via dedicated software clients (including mobile platforms) or via web-browser. Management and control over major physical and virtual cloud infrastructure, including networks, servers and operating systems (except a restricted set of application configuration settings) are executed by cloud provider. The service model can also provide authorized users with access to general-purpose applications and to various specialized systems [8].

**VDI** is the creation of desktops in virtual environment. With the help of desktop virtualization technology, a user having any device with network access (smartphone, tablet, thin client) can receive access to personal desktop and corporate information resources.

**Data remote backup** is a service providing users with system for data backup and storage. Data remote backup systems are embedded into client program. This program capture, crunch, encipher and transfer data to ICP servers.

The fig. 4 below model of user`s interaction with cloud platform.

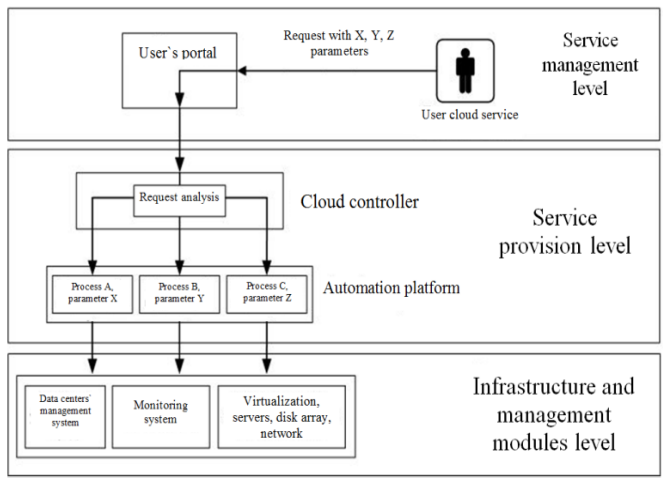

Fig. 4. User`s interaction with cloud platform

Cloud environment user makes a request for service management on service management level (creation, withdrawal, modification, etc.). The request contains the required action and parameters. For example, a user needs to add one more virtual server to computing resource. Such request will require server addition and its parameters will contain the following data: number of CPU, number of RAM, operating system type and so on, including computing cluster identifier to which it is required to add one more node.

The request is made in user-friendly graphics interface and transferred to service provision level to cloud controller. Cloud controller receives and analyses the request. According to analysis result it is required to check availability of required resources, resources reservation for certain user`s needs and afterwards, launch of required processes in underlying automation platform. Automated processes, which are initiated during the scenario implementation, receive required parameters transmitted from user`s request. During execution of the processes, the platform interacts with objects of infrastructure level and management modules. Scenario of virtual server inclusion to virtual cluster will include the following systems: virtualization system for creation of virtual machine with required parameters; servers management system for cluster software installation; platform`s connectors for configuration execution straight on a new node; monitoring system for automatic turn-on of a new node to general monitoring outline.

The fig. 5 demonstrates the logic of cloud platform's interaction with user.

User`s requirements are analyzed and transmitted to the set of automated procedures, which in its turn execute required technical operations. The system`s feedback is initiated by various subsystems of cloud environment on service provision level and infrastructure level. Feedback scenarios implement principles and tasks of computing capacities operation. Interaction between management and monitoring systems and cloud controller represents the main scenarios. Upon execution of some configuration task or receipt of system`s emergency message, the launch of corresponding processes on automation platform is initiated, which in their turn send required requests to cloud controller. Cloud controller associates the requests with services of certain user and publishes information on corresponding home page of service portal. Another example is publication of the results of computing tasks executed in cloud environment.

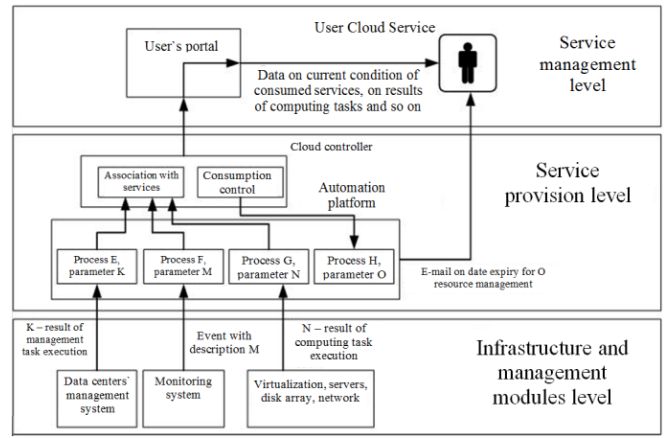

Fig. 5. Cloud platform feedback to user

Another standard scenario is control over the use of computing center services and users timely notification on date expiry via portal or e-mail.

VI. SUBSYSTEM OF RESOURCES VIRTUALIZATION

Subsystem of resources virtualization includes the following modules:

*1) Virtualization of file access*
*2) Virtualization of block access*
*3) Virtualization of data communication networks*
*4) Virtualization of computing resources*
Fig 6 shows a diagram for virtualization subsystem.

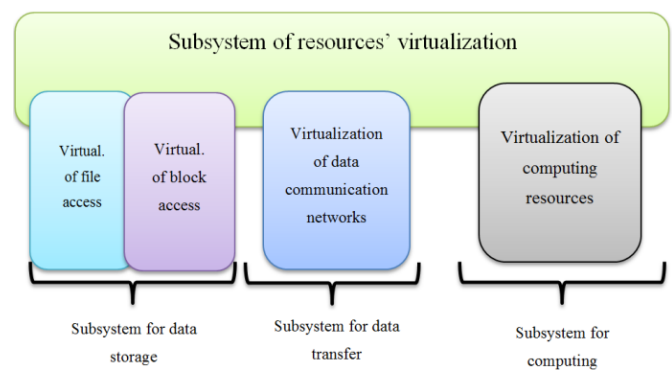

Fig. 6. Block diagram for virtualization subsystem

Virtualization is creation of flexible substitute for physical resources with the same functions and external interface but with different attributes such as size and efficiency. Such substitute is called virtual resources and usually operating systems are not aware of substitution made.

Virtualization is applied to physical hardware via incorporation of several physical resources in one pool from which users can get virtual resources. With the help of virtualization it is possible to create several virtual resources from a physical one.

Virtual resources can have functions or peculiarities that are absent on physical ones. During virtualization, several virtual systems are created from one physical system. Virtual systems are independently operating environments, which use virtual resources.

Usually, system virtualization is executed with the help of hypervisor technology. Hypervisor (irrespective of type) is a multilayer application, which separates hardware from its guest systems. Each guest operating system sees virtual machine instead of physical equipment [9].

### VII. Subsystem of Computing Platform

Subsystem of computing platform is divided on servers designated for solving of centralized management tasks and servers designated for execution of production tasks (service provisions). Combination of management servers is called "management cluster" and combination of managed servers is called "resource group". Only one (resilient) copy of management cluster is installed whereas there might be several resource groups (depends on service levels or equipment set).

It includes the following servers:

- x86 architecture
- with 2 physical processor sockets
- with 4 physical processor sockets
- floor-standing version
- blade version

Main requirements to server equipment:

- availability of build-in technologies for provision of crucial components resiliency;
- duplication of power supply units;
- duplication of I/O interface;
- management of single errors in random access memory;
- selection of compatible processors scale preferably with the similar capacity for consistent operation of software hypervisor;
- for 2 socket servers the number of installed random access memory should be not less than 256 Gb;
- for 4 socket servers the number of installed random access memory should be not less than 512 Gb;
- local HD are not mandatory as servers should download OS from external data storage systems.

### VIII. Conclusion

Application of cloud technologies enables reducing capital and current (operation) costs and increasing the level of equipment use, which amounts to 10% in government sector.

Cloud computing provides a more effective use of government agencies' computing resources and at the same time, the resources are available for all government agencies and can be rationally distributed as workload changes. Balance of government agencies' information and communication assets will be achieved.

Besides, efficiency of computing capacities per kilowatt-hour will increase, which in its turn will lead to rise in environmental friendliness of government agencies performance.

Upon transferring to functions on development of information systems and information assistance with mandatory support of domestic providers to IT-outsourcing, government agencies will be disengaged from non-profile purposes and assets and obtain a qualitative final result with the right for intellectual property use from software developers and service companies; competition in public sector in the field of informatization will grow together with Kazakhstani participation in procurement; budget saving on a nationwide scale will be achieved.

Thus, nowadays, there is a necessity to address the issues through changes in the Republic of Kazakhstan legislation in the field of informatization.

The basic practical relevance of the publication is the list of criteria for selection of cloud platform solution by which optimal cloud products will be chosen for the Republic of Kazakhstan e-Gov considering quality-price ratio, and also the framework of information and communication architecture will be introduced.